\documentclass[11pt,twoside]{article}
\usepackage{asp2010}

\resetcounters

\begin{document}
\title{{\em Spitzer}  Mid-Infrared Photometry of 500 -- 750~K Brown Dwarfs}
\author{S. K. Leggett$^1$, L. Albert$^2$, E. Artigau$^3$, Ben Burningham$^4$, 
X. Delfosse$^5$,\\
P. Delorme$^6$, T. Forveille$^5$, P. W.  Lucas$^4$, M. S. Marley$^7$, 
D. J. Pinfield$^4$, C. Reyl\'e$^8$, D. Saumon$^9$, R. L. Smart$^{10}$, and 
S. J. Warren$^{11}$
\affil{$^1$Gemini Observatory Northern Operations, 670 N. A'ohoku Place,
Hilo, Hawaii 96720, USA }
\affil{$^2$Canada-France-Hawaii Telescope, 65-1238 Mamalahoa Highway, Kamuela, Hawaii 96743, USA }
\affil{$^3$D\'epartement de Physique and Observatoire du Mont M\'egantic, Universit\'e de Montr\'eal, 
C.P. 6128, Succursale Centre-Ville, Montr\'eal, QC H3C 3J7, Canada}
\affil{$^4$Centre for Astrophysics Research, Science and Technology 
Research Institute, University of Hertfordshire, Hatfield AL10 9AB, UK}
\affil{$^5$Laboratoire d'Astrophysique de Grenoble,Universit\'e J. Fourier, CNRS, UMR5571, Grenoble, 
France}
\affil{$^6$School of Physics and Astronomy, University of St Andrews, 
North Haugh, St Andrews KY16 9SS, United Kingdom}
\affil{$^7$NASA Ames Research Center, Mail Stop 245-3, Moffett Field, CA 94035, USA }
\affil{$^8$Besan\c{c}on Observatory, Universite de Franche-Comte, Institut Utinam, UMR CNRS	6213, BP	
1615, 25010 Besan\c{c}on Cedex, France}
\affil{$^9$Los Alamos National Laboratory, P.O. Box 1663, MS F663, Los 
Alamos, \\NM 87545, USA }
\affil{$^{10}$INAF/Osservatorio Astronomico di Torino, Strada 
Osservatorio 20, \\10025 Pino Torinese, Italy }
\affil{$^{11}$Imperial College London, Blackett Laboratory, Prince Consort 
Road, London SW7 2AZ, UK }}

\begin{abstract} Mid-infrared data, including {\it Spitzer} warm-IRAC [3.6] and [4.5] photometry, is 
critical for understanding the cold population of brown dwarfs now being found, objects which have 
more in common with planets than stars. As effective temperature ($T_{\rm eff}$) drops from 800~K to 
400~K, the fraction of flux emitted beyond 3~$\mu$m increases rapidly, from about 40\% to$ >$75\%. This 
rapid increase makes a color like $H$-[4.5] a very sensitive temperature indicator, and it can be 
combined with a gravity- and metallicity-sensitive color like $H-K$ to constrain all three of these 
fundamental properties, which in turn gives us mass and age for these slowly cooling objects.	
Determination of mid-infrared color trends also allows better exploitation of the WISE mission by the 
community. We use new {\it Spitzer} Cycle 6 IRAC photometry, together with published data, to present 
trends 
of color with type for L0 to T10 dwarfs. We also use the atmospheric and evolutionary models of Saumon 
\& Marley to investigate the masses and ages of 13 very late-type T dwarfs, which have $H$-[4.5]$>$3.2 
and $T_{\rm eff} \approx$500~K to 750~K. 

Note: This is an updated version of \citet{legg10a}; a photometry compilation is available at {\it www.gemini.edu/staff/sleggett}. 

\end{abstract}

\section{Introduction}

The last decade has seen a remarkable increase in our knowledge of the bottom of the 
stellar main-sequence and of the low-mass stellar and sub-stellar (brown dwarf) 
population of the solar neighborhood. Two classes have been added to the spectral 
type sequence following M --- L, and T; T dwarfs with effective temperatures ($T_{\rm 
eff}$) as low as 500~K are now known and we are truly finding objects that provide the 
link between the low-mass stars and the giant planets. As $T_{\rm eff}$ decreases, 
brown dwarfs emit significant flux at mid-infrared wavelengths 
\citep[e.g.][]{burr03,legg09}.
At $T_{\rm eff}=$1000~K 30\% of the total flux is emitted at 
wavelengths longer than 3~$\mu$m, while at 600~K 60\% of the flux is emitted in this 
region (according to our models). 

\begin{figure}[!hb]
\plotfiddle{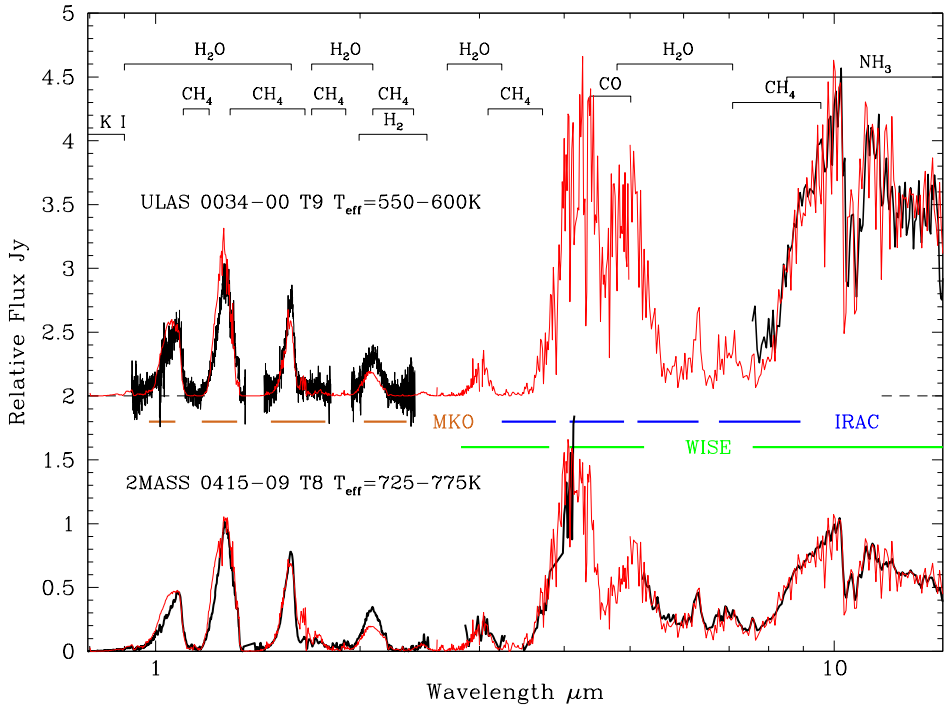}{9cm}{270}{50}{50}{-200}{290}
\caption{Observed (black lines) and modelled (red lines) SEDs of the 750~K T8 dwarf 
2MASS 0415-09 \citep[lower spectrum,][]{saum07},
and the 600~K T9 dwarf ULAS 0034-00 
\citep[upper spectrum,][]{legg09}.
Principal absorption features are identified; 
NH$_3$ is also likely for the T9 dwarf near 1.0, 1.2, 1.3, 1.5 and 1.8~$\mu$m. The MKO-system 
$YJHK$, 
IRAC and WISE filter bandpasses are indicated. }
\end{figure}

Figure 1 shows spectral energy distributions (SEDs) 
for a 750~K T8 dwarf and a 600~K T9 dwarf; the flux emerges from windows between strong 
bands of, primarily, CH$_4$ and H$_2$O absorption. As the temperature drops from 750~K 
to 600~K, the ratio of the mid- to the near-infrared flux increases dramatically, and 
more flux emerges through the windows centered near 5~$\mu$m and 10~$\mu$m. Filter 
passbands are indicated for the near-infrared $YJHK$ MKO-system \citep{toku02},
as well as the {\it Spitzer} IRAC bands \citep{fazi04}
and the three 
shortest-wavelength WISE bands \citep{liu08}.
The IRAC and WISE filters sample 
regions of both high and low flux, thus both cameras are sensitive to cold brown 
dwarfs, which can be identified by extreme colors in their respective photometric systems.

We used {\it Spitzer} Cycle 6 time to obtain IRAC photometry of late-type T 
dwarfs. Here we combine these data with published photometry to examine trends in 
colors with  spectral type, which will be useful for the design and use of ongoing and planned 
infrared surveys. We also examine in detail the colors of the $500 \leq T_{\rm eff}$ K 
$\leq 800$ dwarfs for correlations with the photospheric parameters effective temperature, 
gravity, and metallicity. We find that various colors do provide indicators of 
temperature and gravity, which in turn constrains mass and age when combined with 
evolutionary models.

\section{Trends with Type, and Survey Sensitivities}

\begin{figure}[!hb]
\plotfiddle{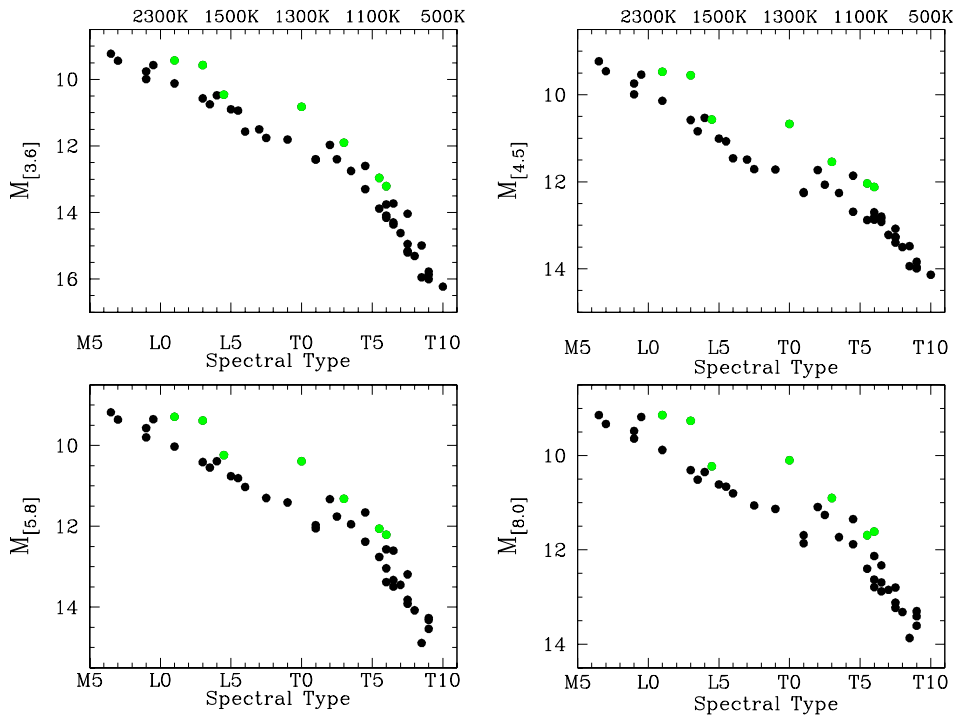}{9cm}{270}{50}{50}{-200}{290}
\caption{Absolute IRAC magnitudes as a function of spectral type; infrared spectral 
type is used for both L and T dwarfs and the uncertainty is 0.5 - 1.0 subclass. Green 
points are known binary systems. $T_{\rm eff}$ values on the top axes 
are from the \citet{step09}
empirical $T_{\rm eff}$:type relationship.
}
\end{figure}

\begin{figure}[!ht]
\plotfiddle{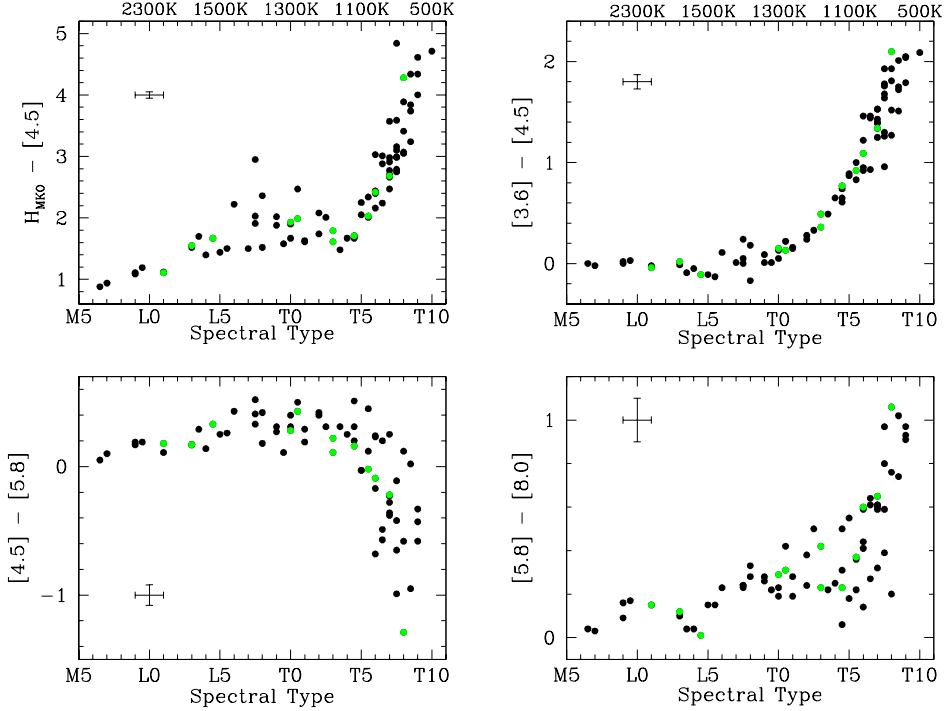}{9cm}{270}{50}{50}{-200}{290}
\caption{ IRAC colors as a function of spectral type; infrared spectral type is used 
for both L and T dwarfs. Green points are known binary systems. Typical uncertainties 
are indicated. $T_{\rm eff}$ values on the top axes 
are from the \citet{step09} empirical 
$T_{\rm eff}$:type relationship.
}
\end{figure}

Figure 2 shows absolute IRAC [3.6], [4.5], [5.8] and [8.0] magnitudes as a function of 
spectral type. For the latest-type T dwarfs, because of the redistribution of the flux 
to the mid-infrared, the drop in intrinsic brightness at these wavelengths is much less 
than in the near- infrared. At $YJHK$ there is an increase of 4 to 5 magnitudes from T5 
to T10, as opposed to the $\sim$2 magnitudes shown in Figure 2. This means that the 
mid-infrared all-sky mission WISE is ideally suited for detecting the very cold brown 
dwarfs. Assuming that the mass function is  flat \citep[e.g.][]{burn10b},
and using the simulations by \citet{burg04},
we estimate that WISE will find 
forty-five 500~K and thirty 400~K brown dwarfs 
\citep[using the sensitivities in][]{main05}.

Figure 3 shows various colors as a function of type. T5 and later types become rapidly 
redder in [3.6]-[4.5] and [5.8]-[8.0], but bluer in [4.5]-[5.8]. The dispersion for 
these types is relatively small compared to the photometric uncertainties. Some scatter 
however is seen in [4.5]-[5.8], most likely due to variations in the CO absorption, 
which impacts the [4.5] flux and which is gravity- and metallicity-sensitive.

\section{Temperature, Metallicity and Gravity}

We now examine observed and modeled trends for the coolest dwarfs in our sample. The 
color $H-$[4.5] is a sensitive indicator of $T_{\rm eff}$ for late-type T dwarfs 
\citep{warr07, step09, legg10a}.
Here, we select 
dwarfs with $H-$[4.5]$ >$ 3.2, which produces a sample cooler than $\sim$800~K. Thirteen 
objects fall into this category, and are listed in Table 1. Figure 4 shows $H-K$ and 
[4.5]-[5.8] as a function of $H$-[4.5], and Figure 5 shows $M_H$ and $M_{[4.5]}$ as a 
function of $H-K$ , $H-$[4.5] and [4.5]-[5.8]. Note that [5.8] photometry is not 
available for dwarfs observed in the post-cryogenic {\it Spitzer} mission. Model sequences 
with a range of gravity and metallicity are also shown \citep{marl02, saum08}.

\begin{figure}[!hb]
\plotfiddle{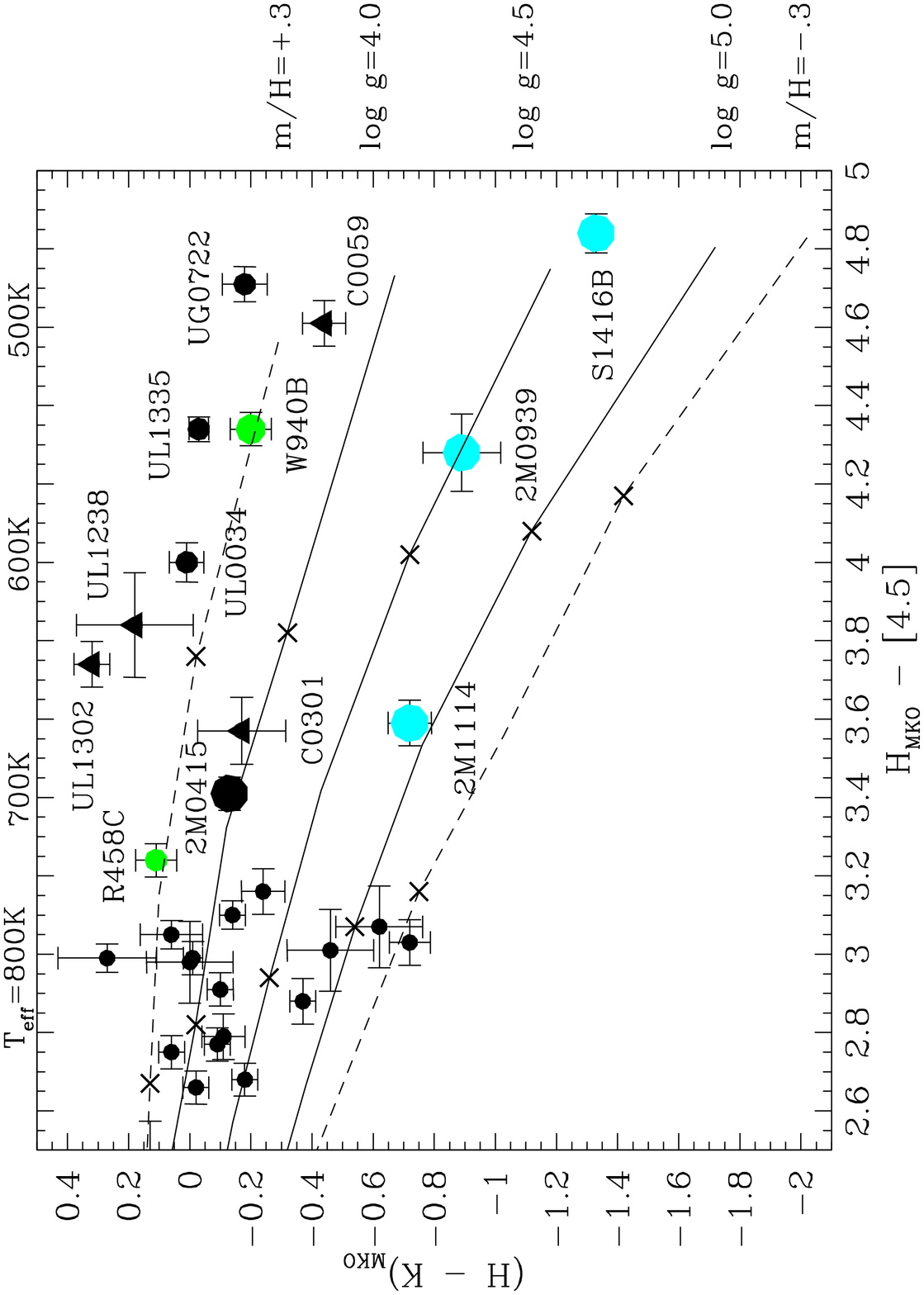}{7cm}{270}{37}{37}{-150}{230}
\plotfiddle{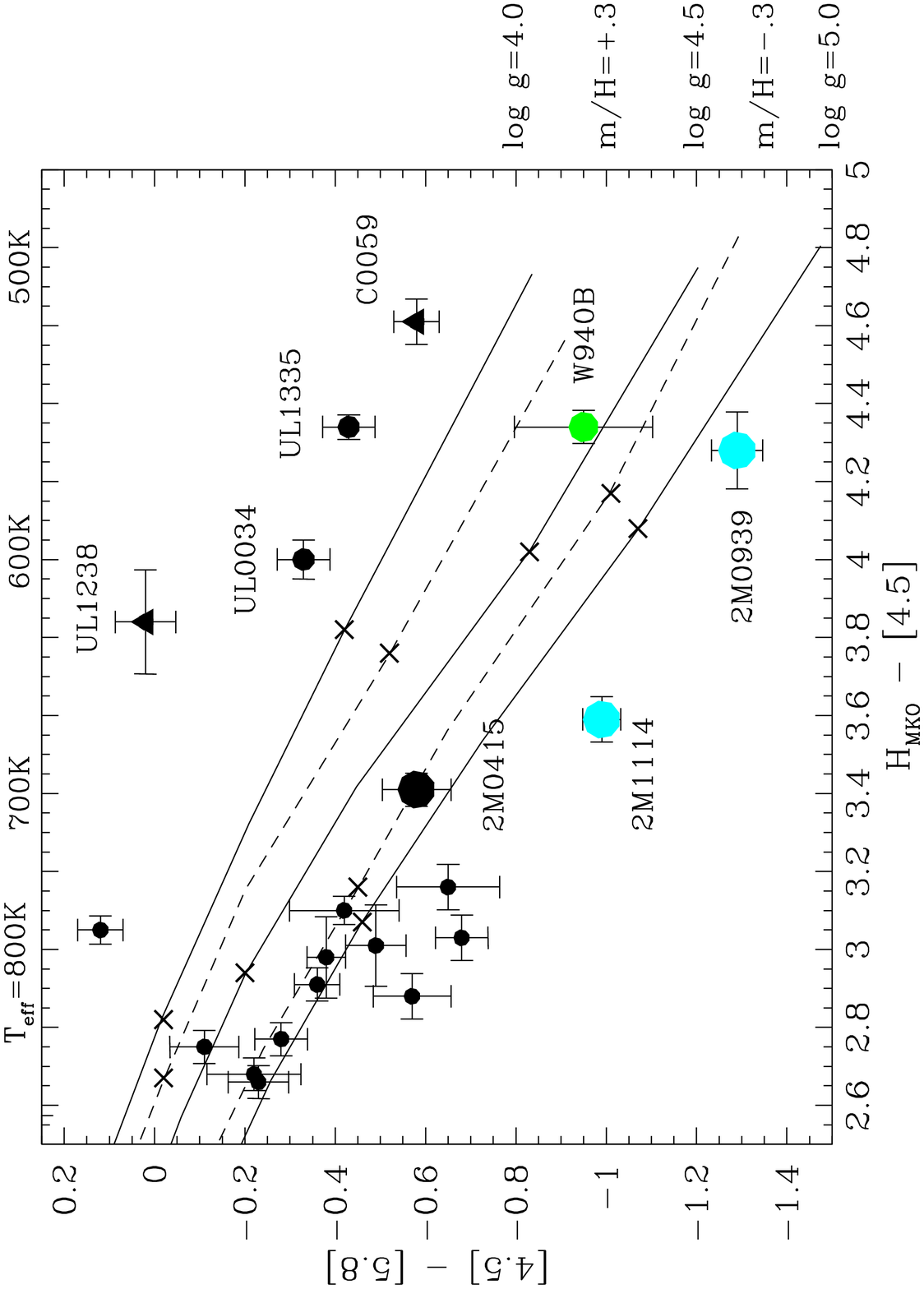}{7cm}{270}{37}{37}{-150}{230}
\caption{ Temperature-sensitive $H-$[4.5]  against gravity- and 
metallicity-sensitive $H-K$ (top) and [4.5]-[5.8] (bottom); dwarfs with 
$H-$[4.5]$>3.2$ are identified and studied in detail. For these objects, symbol 
color indicates metallicity: black is unknown, green is solar, and cyan is metal-poor. 
Size indicates gravity: largest to smallest filled circles have log $ g > 5$, $\sim$ 5.0 
and 
$\sim$ 4.5; triangles have unconstrained gravity.	Model sequences with 
log $g =$ 4.0, 4.5 
and 5.0 and [m/H]$=$0 are shown as solid lines, and log $g=4.5$ with [m/H]$=-0.3$ and 
$+0.3$ 
as 
dashed lines. Crosses indicate the locations along the sequences of the $T_{\rm eff}=$ 
800~K and 
600~K points.
}
\end{figure}

\begin{table}[!ht]
\caption{Sample of Very Late-Type T Dwarfs with $H-$[4.5] $ > 3.2$.}
\smallskip
\begin{center}
{\small
\begin{tabular}{lcccccccc}
\tableline
\noalign{\smallskip}
Name & Spectral & $T_{\rm eff}$ (K) & log $g$ & [m/H] & Mass & Age  & 
\multicolumn{2}{c}{References\tablenotemark{a}} \\
\noalign{\smallskip}
\cline{8-9}
\noalign{\smallskip}
            &    Type       &                           &                &    & (Jupiter)                          
&       (Gyr)            & A &  B\\
\noalign{\smallskip}
\tableline
\noalign{\smallskip}
CFBDS 0301-16 & T7 & 700 - 750 & $\sim$5.0 & $\sim$0.0 & 20 - 40 & 0.4 - 4.0 & 1 & 2 \\
2MASS 1114-26 & T7.5 & 725 - 775 & 5.0 - 5.3 & -0.3 & 30 - 50 & 3 - 8 & 3 & 4 \\
SDSS 1416+13B & T7.5 & 500 - 700 & 4.8 - 5.6 & $\leq -0.3$ & 20 - 45 & 2 - 10 & 5, 6 & 5, 7 \\
2MASS 0415-09 & T8 & 725 - 775 & 5.0 - 5.4 & $\geq 0$ & 33 - 58 & 3 - 10 & 8 & 9 \\
2MASS 0939-24 & T8 & 500 - 700 & 5.0 - 5.3 & $-0.3$ & 20 - 40 & 2 - 10 & 3 & 10 \\
Ross 458C & T8.5 & 700 - 750 & $\sim 4.5$ & 0.0 & 7 - 20 & 0.1 - 1.0 & 11 & 2 \\
Wolf 940B & T8.5 & 585 - 625 & 4.8 - 5.2 & 0.1  & 24 - 45 & 3 - 10 & 12 & 13 \\
ULAS 1302+13 & T8.5 & 650 - 700 & 4.0 - 4.5 & $>0$ & 5 - 15 & 0.1 - 0.4 & 14 & 2 \\
ULAS 1238+09 & T8.5 & 575 - 625 & 4.0 - 4.5 & $\geq 0$ & 6 - 10 & 0.2 - 1.0 & 15 & 16 \\
CFBDS 0059-01 & T9 & 500 - 550 & 4.5 - 5.0 & $\sim 0$ & 10 - 30 & 1 - 10 & 17 & 2 \\
ULAS 0034-00 & T9 & 550 - 600 & 4.5  & $\geq 0$ & 13 - 20 & 1 - 2 & 18 & 19 \\
ULAS 1335+11 & T9 & 500 - 550 & 4.5  & $\geq 0$ & 5 - 20 & 0.1 - 2 & 15 & 10 \\
UGPS 0722-05 & T10 & 480 - 560 & 4.0 - 4.5  & $\sim 0$ & 5 - 15 & 0.2 - 2.0 & 20 & 20 \\
\noalign{\smallskip}
\tableline
\noalign{\smallskip}
\multicolumn{9}{p{1.05\textwidth}}
{$^a$ References are to discovery (A) and parameter derivation (B).
(1) \citet{reyl10};
(2) this work;
(3) \citet{tinn05}; 
(4) \citet{legg07}; 
(5) \citet{burn10a};
(6) \citet{scho10}; 
(7) \citet{burg10}; 
(8) \citet{burg02}; 
(9) \citet{saum07}; 
(10) \citet{legg09}; 
(11) \citet{gold10}; 
(12) \citet{burn09}; 
(13) \citet{legg10b}; 
(14) \citet{burn10b}; 
(15) \citet{burn08};
(16) \citet{legg10a};
(17) \citet{delo08};
(18) \citet{warr07}; 
(19) \citet{smar10}; 
(20) \citet{luca10}.}
\end{tabular}
}
\end{center}
\end{table}

Some of these objects are well-studied, and we indicate in the figures metallicity and 
gravity where these are known; these and other parameters are listed in Table 1. It 
can be seen that, while there is not exact agreement between models and data, the 
general trends are reproduced, with metal-poor and high-gravity dwarfs being blue in 
$H-K$ and [4.5]-[5.8], and slightly redder in $H-$[4.5]. The $H-K$ color is impacted by 
pressure-induced H$_2$ opacity which is sensitive to metallicity and to a lesser extent 
gravity. The [4.5] flux is impacted by a gravity- and metallicity-sensitive CO band. 
There is degeneracy between the effects of gravity and metallicity, but there is an 
indication that $H-K$ is more sensitive to metallicity and [4.5]-[5.8] to gravity, as 
the models calculate. The location of the dwarfs in Figures 4 and 5 relative to 
benchmark dwarfs, allows temperature, metallicity and gravity to be newly estimated for 
four dwarfs, and these parameters are given in Table 1. Note that these values do not 
rely on the absolute values implied by the models, but instead the models are used in a 
relative sense, using well-studied dwarfs as benchmarks.

\begin{figure}[!ht]
\plotfiddle{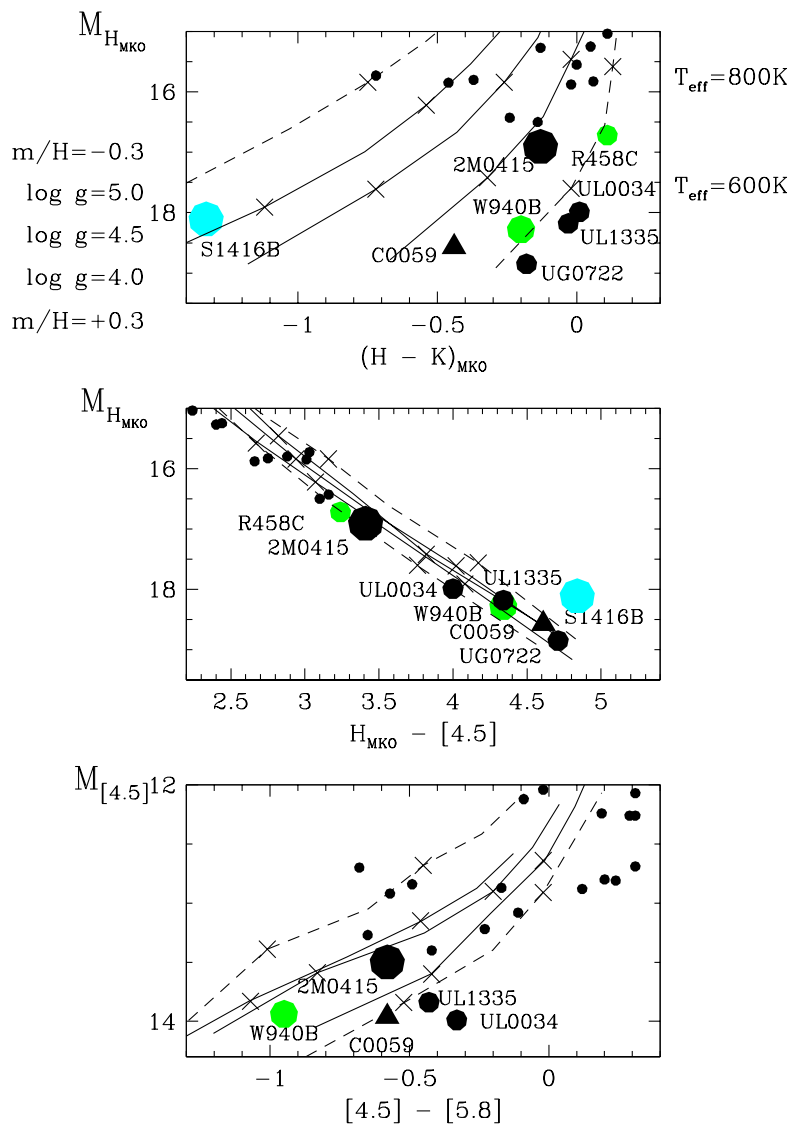}{15cm}{0}{60}{60}{-200}{-30}
\caption{ 
Absolute $H$ and [4.5] magnitudes as a function of various colors. Symbols and line types are as in 
Figure 4.}
\end{figure}

\section{Conclusions}

The IRAC data implies that the majority of the UKIDSS 500~K to 600~K dwarfs are young and 
low mass, a result not currently understood in terms of selection effects or the mass 
function. Photometric data at wavelengths longer than 3~$\mu$m are both important and 
useful for the latest-type T dwarfs with $500 \leq T_{\rm eff}$ K $\leq 800$, and will 
be even more so for the cooler objects expected to be found by WISE and other sky 
surveys. It is only possible to get such data from the ground for very bright objects 
and mid-infrared space missions are crucial for continued progress in this field.

\acknowledgements 

This work is based on observations made with the {\it Spitzer} Space Telescope, which is operated by the Jet Propulsion Laboratory, California Institute of Technology under a contract with NASA. Support for this work was provided by NASA through an award issued by JPL/Caltech. Support for this work was also provided by the {\it Spitzer} Space Telescope Theoretical Research Program, through NASA. SKL's research is supported by the Gemini Observatory, which is operated by the Association of Universities for Research in Astronomy, Inc., on behalf of the international Gemini partnership of Argentina, Australia, Brazil, Canada, Chile, the United Kingdom, and the United States of America.

\bibliography{Leggett_S}

\end{document}